# Fermilab's Transition to Token Authentication

*Dave* Dykstra[1*], *Mine* Altunay[2], *Shreyas* Bhat[1], *Dmitry* Litvintsev[1], Marco Mambelli[1], Marc Mengel[1], *Stephen* White[1]

[1]Scientific Computing Systems and Services Division, Fermilab, Batavia, IL, USA
[2]Security and Emergency Management Division, Fermilab, Batavia, IL, USA

**Abstract.** Fermilab is the first High Energy Physics institution to transition from X.509 user certificates to authentication tokens in production systems. All the experiments that Fermilab hosts are now using JSON Web Token (JWT) access tokens in their grid jobs. Many software components have been either updated or created for this transition, and most of the software is available to others as open source. The tokens are defined using the WLCG Common JWT Profile. Token attributes for all the tokens are stored in the Fermilab FERRY system which generates the configuration for the CILogon token issuer. High security-value refresh tokens are stored in Hashicorp Vault configured by htvault-config, and JWT access tokens are requested by the htgettoken client through its integration with HTCondor. The Fermilab job submission system jobsub was redesigned to be a lightweight wrapper around HTCondor. The grid workload management system GlideinWMS which is also based on HTCondor was updated to use different tokens for pilot job submission and in-framework authentication. For automated job submissions a managed tokens service was created to reduce duplication of effort and knowledge of how to securely keep tokens active. The existing Fermilab file transfer tool ifdh was updated to work seamlessly with tokens, as well as the Fermilab POMS (Production Operations Management System) which is used to manage automatic job submission and the RCDS (Rapid Code Distribution System) which is used to distribute analysis code via the CernVM FileSystem. The dCache storage system was reconfigured to accept tokens for authentication in place of X.509 proxy certificates. As some services and sites have not yet implemented token support, proxy certificates are still sent with jobs for backwards compatibility, but some experiments are beginning to transition to stop using them. There have been some glitches and learning curve issues but in general the system has been performing well and is being improved as operational problems are addressed.

[*] Corresponding author: dwd@fnal.gov

## 1 Introduction

Since its inception, the Worldwide LHC Computing Grid (WLCG) has used X.509 proxy certificate user credentials for authentication and authorization of grid jobs. That technique never came into common use in industry and is becoming increasingly difficult to support, so the WLCG community concluded that it was time to replace them. The two primary industry standards chosen as a replacement were JSON Web Tokens [1] (JWTs) and OAuth 2.0 [2], including the OpenID Connect (OIDC) [3] identity layer. JWTs are verifiable offline, which is a very important quality (shared by X.509 proxy certificates) for them to be usable on the grid scale. JWTs can also be very fine-grained, which is potentially more secure than X.509 proxy certificates because they can be generated with fewer permissions.

Fermilab has been using tokens based on the WLCG Common JWT Profile [4] for authorization of all grid jobs initiated there since early 2023. This paper provides a brief technical overview of the major components that were either updated or created for that transition. These are the components numbered by the paper section where they are discussed:

2. The Frontier Experiments RegistRY (FERRY) [5] contains the database that keeps track of what authorizations each user has and makes that available to the token issuer CILogon [6].
3. High security-value refresh tokens are stored in a Hashicorp Vault [7] service that is configured by an htvault-config [8] package, accessed with its client htgettoken [9, 10], and integrated with HTCondor [11].
4. The Fermilab job submission system jobsub [12] is a lightweight wrapper around HTCondor.
5. The Managed Tokens service [13, 14] makes sure that tokens that are used for unattended "robot" operations stay refreshed in HTCondor.
6. GlideinWMS [15] is the grid job Workload Management System that is also based on HTCondor.
7. Data storage is managed by dCache [16].
8. Other components updated for tokens were the data handling tool ifdh [17], the Production Operation Management System (POMS) [18] which manages large job batches, and the Rapid Code Distribution System (RCDS) [19] which quickly distributes experimenter software via the CernVM FileSystem [20].

## 2 FERRY & CILogon

FERRY [5] is Fermilab's grid access control and quota management service. It was written within the last decade to manage information about collaborators on the experiments hosted at Fermilab that use grid computing resources. It was extended for this transition to manage all the information needed to generate tokens and to translate that information into an LDAP server hosted by the token issuer CILogon [6]. The format of that LDAP data was negotiated between developers at Fermilab and CILogon.

The LDAP data is primarily divided into two parts. The first part lists each collaborator and which experiments and roles within the experiments they are authorized for. The second part lists all the token scopes that are associated with each role.

FERRY has a web API for updating and reading information. The API was extended for this project to be able to accept CILogon-issued JWTs to authenticate access.

Currently Fermilab hosts on the order of 30 experiments, some with a large number of collaborators and some with just a small number. Five of the larger experiments have been set up to have a token issuer with CILogon matching their name, but the rest all share one called “fermilab”. The smaller experiments are each assigned storage in the shared area beginning with their name, and the tokens also indicate a group name that they are part of so that’s how they are kept separate.

## 3 Vault, htvault-config, htgettoken, and HTCondor integration

The next major piece that was identified as a necessary component for managing tokens was an OAuth client that worked well with Linux command-line grid job submission. No adequate client was found to be in existence, so that component needed to be created. In order to balance the needs of security and convenience, it was decided to introduce another service, with the OAuth client on a shared secured server and a command line client for that server. Hashicorp Vault [7] already had almost all the capabilities needed, so it was chosen as the server. The htvault-config [8] package was created to configure Vault, and the Vault client htgettoken [9, 10] was also created.

The first time someone tries to obtain a token for a given experiment and role the request is redirected to go through OIDC authentication in a web browser. When successful, that returns a high security-value refresh token. It has high security value because it lasts for 4 weeks and is infinitely renewable. That refresh token stays in Vault, and instead Vault generates its own token that lasts for only 7 days. After that token expires, a new 7-day token can be obtained without a web browser, using either Kerberos or ssh-agent authentication.

Vault tokens authorize requests to Vault to use a stored refresh token to obtain a JWT access token from the token issuer. Access tokens are configured to be very short-lived, 3 hours, because they are widely distributed in jobs and therefore most vulnerable to being stolen. The time was not made shorter because some time is needed to intervene if one of the services develops a problem. Because the tokens are short-lived, mechanisms to refresh them are needed. The HTCondor [11] integration is a vital part of refreshing the tokens in running jobs.

### 3.1 htvault-config

The htvault-config package was created to make it easy to configure Vault for this purpose. Often the combination of the htvault-config package and Vault are referred to as just “HTVault”. Configuration is done through simple yaml files. Changes can be made to the yaml files on a running system, and only the differences are applied. The package includes some Vault plugins that are applied on top of the standard vault package. It includes options for configuring some valuable core Vault functions, including client rate limiting and High Availability using 3 servers.

There is also a Fermilab-specific script called htvault-gen that generates most of the htvault-config yaml files from data in FERRY. For that reason the HTVault operators do not need to do anything when new experiments or new roles are added; the new configuration is automatically sent to both HTVault and CILogon. Fermilab uses Kerberos, so HTVault ssh-agent authentication is not configured there.

Separate long-lived OAuth credentials in the form of client IDs and secrets are configured in HTVault for each of the six defined CILogon token issuers. The smaller experiments that share the same “fermilab” token issuer at CILogon are each configured with their own name in HTVault, so to the HTVault client they appear to have their own token issuer.

### 3.2 htgettoken

The command line client for HTVault is called htgettoken. Its purpose is to automate the flows through the various HTVault API endpoints. End users usually don't need to directly invoke htgettoken because higher level scripts do that for them.

These additional useful tools are part of the htgettoken package:

1. htdecodetoken – displays the JSON contents of the JWT token. The source of the token is located by following the WLCG Bearer Token Discovery [21] standard.
2. htdestroytoken – deletes both an access token and a vault token, if present.
3. httokensh – provides an access token to a command and keeps renewing the token as needed as long as the command runs.

### 3.3 HTCondor integration

An integration with HTVault was added to HTCondor, so that when a job is submitted the necessary tokens will be automatically obtained and access tokens automatically refreshed in running jobs.

The package containing the integration is called condor-credmon-vault. On the job submission machine, the package configures HTCondor to have the condor_submit command call out to a script called condor_vault_storer. On the schedd machine, the condor-credmon-vault package adds a program called condor_credmon_vault as a plugin to condor_credd. When condor_vault_storer first runs it obtains a 4-week vault token and stores it into condor_credmon_vault, where condor_credd uses it regularly to keep access tokens refreshed in jobs. condor_vault_storer does not store the 4-week vault token on the submit machine but instead exchanges it for a 1-week vault token to store there. Figure 1 shows these components in the context of a full HTCondor deployment and shows the places that the different types of tokens are transferred.

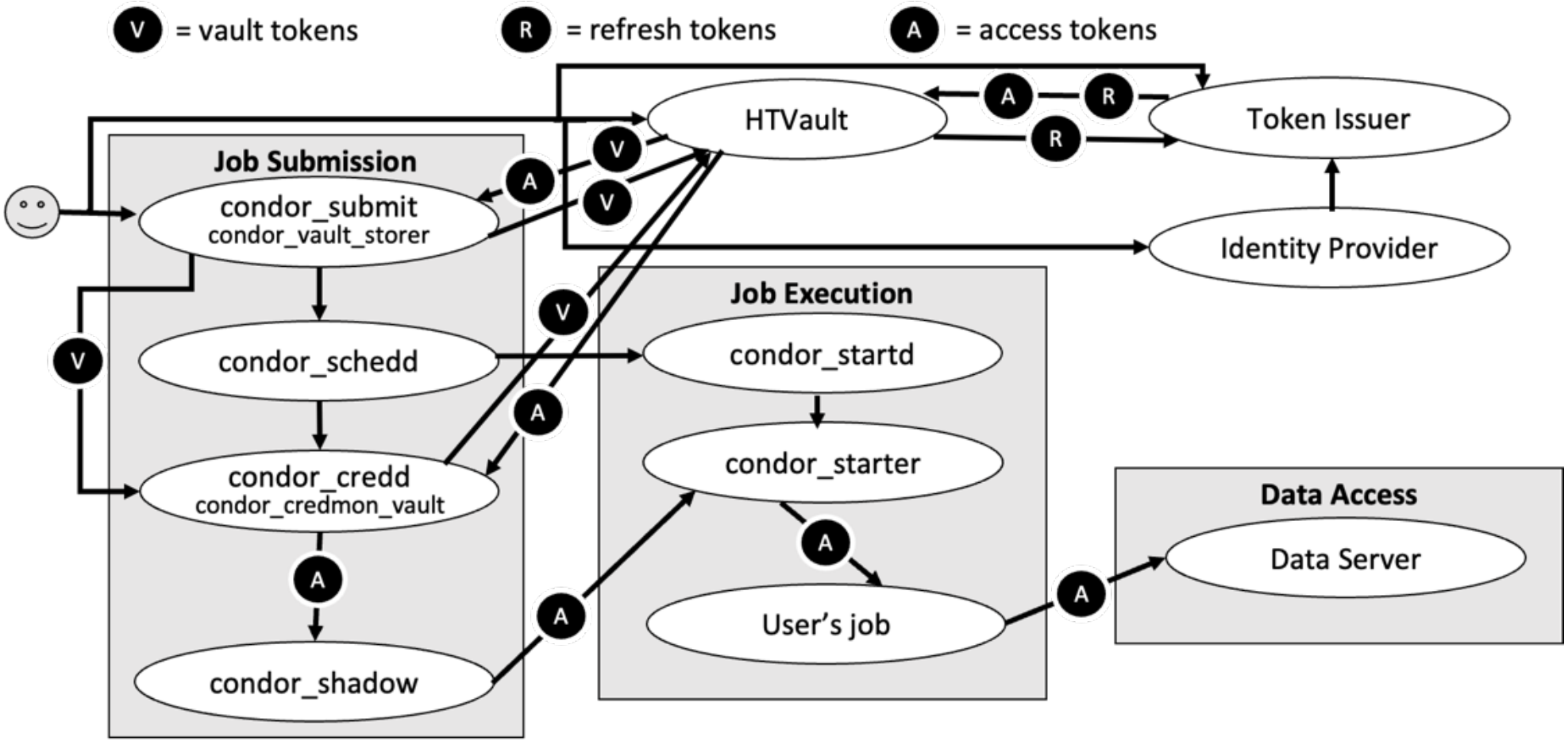

**Figure 1.** HTVault integration components in a full HTCondor deployment

HTCondor already had a mechanism to specify scopes and/or the audience (i.e. intended recipients) of OAuth-created tokens in job submit files and send the tokens to jobs with given names. The HTVault integration changed that only slightly to first use tokens with a default list of scopes based on named roles and then optionally downgrade them to weaker tokens with fewer scopes and/or restricted audiences. All the information regarding the token is sent to condor_vault_storer which takes care of storing it in condor_credd. From there condor_credmon_vault contacts HTVault as needed to fetch all the tokens needed to refresh in jobs.

## 4 jobsub

The Fermilab job submission system jobsub [12] was originally designed with its own client server model, with most of the logic on the server side. That approach got to be unwieldy over time, and meanwhile the capabilities of HTCondor were enhanced, so it was decided to rewrite the system from scratch as a new system sometimes known as jobsub_lite and sometimes known as just the new version of jobsub. Instead of having its own server, jobsub_lite was to be only a lightweight wrapper around the condor client tools while still being largely compatible with the old jobsub command line interface. Support for tokens was included in the design of jobsub_lite from its beginning.

jobsub_lite as used at Fermilab is configured to be able to submit jobs to multiple different HTCondor schedd servers, and to use the same tokens that are used in jobs to authorize the clients to those servers. It also supports many different experiments, including enabling individuals to submit jobs for multiple experiments. It keeps track of separate tokens for each schedd and experiment.

jobsub_lite generates the job submission files accepted by condor_submit, so command line options were added for specifying reduced scopes and audiences in tokens. A full set of options for job submission frequently gets quite long, so typically experts on each experiment write scripts to make it easy for other collaborators to submit jobs.

## 5 Managed Tokens service

The components discussed above are sufficient to securely manage tokens for interactive use, but there are also many use cases where jobs need to be submitted by automated scripts, often referred to as "robots". Such cases need to use long-lived credentials so they can keep running without interventions. To keep those long-lived credentials from needing to be exposed on many different machines for different experiments, and to avoid duplicating the knowledge and code for managing the credentials in many places, a Managed Tokens [13] service was created.

When a new robot is added, the operator of the Managed Tokens service uses OIDC authentication to get the process started. The operator also obtains a long-lived Kerberos credential corresponding to the token made for that robot, so the service can renew a vault token when the old one expires. The service then directly calls condor_vault_storer for each schedd used by the experiment associated with the robot, so each credd is always kept up to date with vault tokens. Then the service copies the vault token onto the robot's machine to keep it updated, using rsync over ssh and the long-lived Kerberos credential for authentication.

For more details on the Managed Tokens service see the paper in these proceedings [14].

## 6 GlideinWMS

GlideinWMS [15] is the Workload Management System that Fermilab uses to manage submitting pilot jobs to grid sites. It submits pilot jobs to remote Computing Elements (CE). Once pilot jobs have started running in a batch slot they call back to GlideinWMS to get payload jobs to run.

With X.509, GlideinWMS was using a single proxy certificate for different functions: to authenticate to computing resources to submit pilot jobs, to enable access to VO services for all user jobs handled by a pilot, and to secure communication within its framework. For transitioning to tokens, GlideinWMS had to revise its credentials infrastructure to be able to support multiple credentials and a finer temporal and spatial granularity [22]. For user credentials it is taking advantage of HTCondor token support, but CE and VO tokens are managed by the experiments via added flexible plug-ins. Some experiments use the Managed Token service to get credentials but, since the machines involved are already kept secure, some use simpler tools like the osg-token-renewer package [23] based on oidc-agent [24]. Either way the credentials are then used via HTCondor to authenticate with the resources or forwarded securely to the pilot jobs to be available for VO services. Finally, the infrastructure tokens are compatible with HTCondor IDTOKENs, and GlideinWMS generates them dynamically for the resources currently provisioned and forwards them securely to the pilots so they can join the virtual cluster and run payload jobs.

## 7 dCache

The software package used to manage experiment data at Fermilab is dCache [16]. The developers of dCache actively participated in defining the WLCG Common JWT Profile, so to transition it to use tokens was primarily a matter of configuring it and working through bug fixes; there was no need for any additional major software development.

## 8 ifdh, POMS, and RCDS

There were also a few other existing grid software components that needed upgrades for the transition to token authentication.

### 8.1 ifdh

Fermilab's data handling command line tool ifdh [17] (named for intensity frontier data handler) is a front end to several other tools used for managing data. It was extended for the token transition to manage tokens in much the same way that jobsub does, except without storing any tokens in HTCondor servers. It invokes htgettoken to automatically get tokens when needed and stores them in the same places as jobsub so both tools can share those tokens.

### 8.2 POMS

Fermilab's Production Operation Management System (POMS) [18] is a web-based tool for managing large batches of job submissions. It fits the definition of "robot", and for production activities, it uses vault tokens distributed by the Managed Tokens service. However, for analysis users, it had some unique requirements, and for that use case, it didn't fit in well as a customer of the Managed Tokens service. For the analysis users' case, as POMS sits on a secured machine, it was given its own long-lived credentials and needed to reuse some of the logic used in the Managed Tokens service, including invoking condor_vault_storer in similar ways.

### 8.2 RCDS

The Rapid Code Distribution System (RCDS) is Fermilab's name for its installation of the cvmfs-user-pub [18] package. The package enables on-demand publication of software into CVMFS [20] via an authenticated web API. Jobsub has an option for users to pass in a tarball of files to publish at the same time as the jobs that use those files. cvmfs-user-pub was updated to accept publication requests from any request with a token containing a "compute.create" scope from a configured list of token issuers.

## 9 Conclusions

Authentication is a core functionality needed for distributed computing, so transitioning it to new technology was a very large undertaking affecting many components. Even so, Fermilab has successfully made that transition and the system for the most part has been functioning very well. Some workflows are still using X.509 proxy user certificates in addition to tokens as of this writing, but that is all scheduled to be turned off by the end of April 2025.

Most of the software referenced in this paper is available as open source for use by others. The Hashicorp Vault license was changed since this project began to not be fully open source, to include a non-compete clause that does not impact Fermilab. However, an open source fork of the project called OpenBao [25] is under active development and has been demonstrated to work as a replacement. Fermilab expects to switch to use that fork in the future.

## Acknowledgments

The authors' work was performed using the resources of the Fermi National Accelerator Laboratory (Fermilab), a U.S. Department of Energy, Office of Science, HEP User Facility. Fermilab is managed by Fermi Forward Discovery Group, LLC, acting under Contract No. 89243024CSC000002.